\documentclass[12pt]{article}
\usepackage{amsmath}
\usepackage{graphicx}
\usepackage{subfigure}
\usepackage{hyperref}

\begin{document}

\author{Ernst Trojan and George V. Vlasov \and \textit{Moscow Institute of Physics
and Technology}}
\title{Tachyonic thermal excitations and causality}
\maketitle

\begin{abstract}
We consider an ideal Fermi gas of tachyonic thermal excitations as a
continuous medium and establish when it satisfies the causality condition.
At high temperature the sound speed is always subluminal $c_s<1$, but there
is no stable form of tachyon matter below the critical temperature $%
T<T_c=0.23m$ that depends on the tachyon mass $m$. The pressure $P$ and
energy density $E$ cannot be arbitrary small, but $P$ can exceed $E$, and $%
P=2.36E$ when $T\rightarrow T_c$.
\end{abstract}


\section{Introduction}

Tachyons, first introduced for the description of superluminal motion \cite
{BDS62,F67}, are commonly known in the field theory as instabilities with
energy spectrum 
\begin{equation}
\varepsilon _k=\sqrt{k^2-m^2}\qquad k>m  \label{t}
\end{equation}
where $m$ is the tachyon mass and relativistic units $c=\hbar =1$\ are used.
The concept of tachyon fields plays significant role in the modern research,
and tachyons are considered as candidates for the dark matter and dark
energy \cite{D1,D2}, they often appear in brane theories \cite{S1998} and
cosmological models \cite{FKS02,S2006}.

A system of many tachyons can be studied in the frames of statistical
mechanics \cite{M84,DHR89}, and thermodynamical functions of ideal tachyon
Fermi and Bose gases are calculated \cite{KRS07,KRS07b}. We have recently
studied the equation of state (EOS) and acoustic properties of the cold
tachyon Fermi gas, which cannot be stable at arbitrary low density \cite
{TV2011b}.

In the present paper we consider fermionic thermal excitations that can
appear in material medium at finite temperature. Thermal excitations with
the energy spectrum 
\begin{equation}
\varepsilon _k=\sqrt{k^2+m^2}  \label{br}
\end{equation}
and finite energy gap $m=\Delta $ are well known in the theory of
superconductivity \cite{CG95,O2002}. Excitations with tachyonic energy
spectrum (\ref{t}) are something exotic. We consider an ideal gas of thermal
excitations as a continuous medium, calculate the sound speed 
\begin{equation}
c_s^2=\frac{dP}{dE}  \label{s}
\end{equation}
and check the causality condition 
\begin{equation}
c_s\leq 1  \label{c}
\end{equation}
The main task of the study: to clarify either (\ref{c}) is satisfied at
arbitrary temperature $T$ or only within narrow range of thermodynamical
parameters.

\section{Tachyon Fermi gas}

Consider a system of free fermions with the energy spectrum $\varepsilon _k$%
. Its energy density $E$, pressure $P$ and particle number density $n$ are
defined by formuals~\cite{Kapusta89}: 
\begin{equation}
E=\frac \gamma {2\pi ^2}\int \varepsilon _k\,f_k\,k^2dk  \label{e}
\end{equation}

\begin{equation}
P=\frac \gamma {6\pi ^2}\int \frac{\partial \varepsilon _k}{\partial k}%
f_k\,k^3dk  \label{p}
\end{equation}
\begin{equation}
n=\frac \gamma {2\pi ^2}\int f_k\,k^2dk  \label{n}
\end{equation}
where $\gamma $ is the degeneracy factor, and 
\begin{equation}
f_k=\frac 1{\exp \left( \varepsilon _k-\mu \right) +1}  \label{f}
\end{equation}
is the distribution function. If we consider thermal excitations in some
material medium, their number will not be conserved, and we put the chemical
potential equal to zero $\mu =0$.

For massive subluminal excitations with the energy spectrum (\ref{br}) the
thermodynamical functions (\ref{e})-(\ref{n}) are determined by formulas 
\begin{equation}
E=\frac{\gamma T^4}{2\pi ^2}\int\limits_\beta ^\infty \frac{\sqrt{x^2-\beta
^2}x^2dx}{\exp x+1}  \label{e2}
\end{equation}

\begin{equation}
P=\frac{\gamma T^4}{6\pi ^2}\int\limits_\beta ^\infty \frac{\sqrt{\left(
x^2-\beta ^2\right) ^3}dx}{\exp x+1}  \label{p2}
\end{equation}
\begin{equation}
n=\frac{\gamma T^3}{2\pi ^2}\int\limits_\beta ^\infty \frac{\sqrt{x^2-\beta
^2}xdx}{\exp x+1}  \label{n2}
\end{equation}
where we have introduced dimensionless variables 
\begin{equation}
x=\frac{\varepsilon _k\,}T\qquad \beta =\frac mT  \label{dim}
\end{equation}

For excitations with the tachyonic energy spectrum (\ref{t}) the formulas (%
\ref{e})-(\ref{n}) are presented so

\begin{equation}
E=\frac{\gamma T^4}{2\pi ^2}\int\limits_0^\infty \frac{\sqrt{x^2+\beta ^2}%
x^2dx}{\exp x+1}  \label{e1}
\end{equation}

\begin{equation}
P=\frac{\gamma T^4}{6\pi ^2}\int\limits_0^\infty \frac{\sqrt{\left(
x^2+\beta ^2\right) ^3}dx}{\exp x+1}  \label{p1}
\end{equation}
\begin{equation}
n=\frac{\gamma T^3}{2\pi ^2}\int\limits_0^\infty \frac{\sqrt{x^2+\beta ^2}xdx%
}{\exp x+1}  \label{n1}
\end{equation}

The thermodynamical functions of massless fermionic excitations will be
given by the same formulas (\ref{e1})-(\ref{n1}) or (\ref{e2})-(\ref{n2})
when we put $\beta =0$ in all integrals.

The particle number density of tachyons and massive subluminal particles is
given in Fig.~\ref{h0}. The ratio $P/E$ vs $\beta =m/T$ is given in Fig.~\ref
{h1}. The equation of state of ultrarelativistic gas 
\begin{equation}
P=\frac E3  \label{photon}
\end{equation}
is achieved in the high-temperature limit $\beta \rightarrow 0$.

\begin{figure}[tbp]
\caption{Sound speed $c_s^2$ vs $\beta = m/T$. Solid line: tachyonic
fermions, dotted line: massive subluminal fermions }
\label{h2}{\includegraphics[scale=0.6]{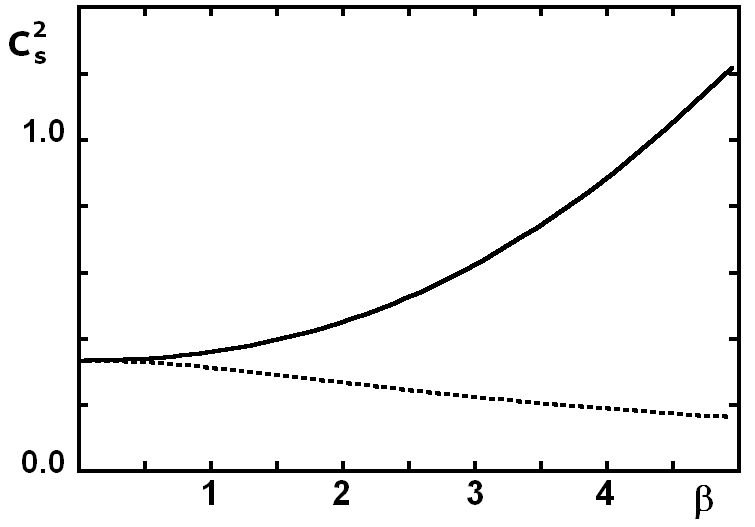}}
\end{figure}

Substituting (\ref{e1}) and (\ref{p1}) in (\ref{s}) we find the sound speed 
\begin{equation}
c_s^2=\frac{dP}{dT}\left( \frac{dE}{dT}\right) ^{-1}  \label{s1}
\end{equation}
that is 
\begin{equation}
c_s^2=\frac{\int\limits_0^\infty \beta ^2\sqrt{x^2+\beta ^2}-\frac 43\sqrt{%
\left( x^2+\beta ^2\right) ^3}\,\frac{dx}{e^x+1}}{\int_0^\infty \left( \frac{%
\beta ^2}{\sqrt{x^2+\beta ^2}}\,-4\sqrt{x^2+\beta ^2}\right) x^2\,\frac{dx}{%
e^x+1}}  \label{s2}
\end{equation}
The sound speed in the gas of massive subluminal fermions is calculated by
formula (\ref{s1}) with the energy density and pressure taken from (\ref{e2}%
) and (\ref{p2}). The behavior of the sound speed is explained in Fig.~\ref
{h2}. At high temperature ($\beta \ll 1$) it tends to the limit $c_s=1/\sqrt{%
3}$ corresponding to the sound speed in ultrarelativistic ideal gas.

The heat capacity is determined by formula 
\begin{equation}
C_V=T\frac{\partial S}{\partial T}=-\beta \frac{\partial S}{\partial \beta }
\label{cv0}
\end{equation}
where the entropy density is 
\begin{equation}
S=\frac{\partial P}{\partial T}=-\frac{\beta ^2}m\frac{\partial P}{\partial
\beta }  \label{ss}
\end{equation}
Substituting (\ref{p1}) in (\ref{cv0}) and in (\ref{ss}), we have 
\begin{equation}
C_V=\frac{\gamma T^3}{2\pi ^2}\int\limits_0^\infty \left( 4\sqrt{\left(
x^2+\beta ^2\right) ^3}-5\beta ^2\sqrt{x^2+\beta ^2}+\frac{\beta ^4}{\sqrt{%
x^2+\beta ^2}}\right) \,\frac{dx}{e^x+1}\,  \label{cv}
\end{equation}
It is always larger than the heat capacity of massless Fermi gas 
\begin{equation}
C_0=\frac{7\gamma \pi ^2T^3}{60}  \label{c0}
\end{equation}
and the latter is larger than the heat capacity of massive subluminal
fermions 
\begin{equation}
C_V=\frac{\gamma T^3}{2\pi ^2}\int\limits_\beta ^\infty \left( 4\sqrt{\left(
x^2-\beta ^2\right) ^3}+5\beta ^2\sqrt{x^2-\beta ^2}+\frac{\beta ^4}{\sqrt{%
x^2-\beta ^2}}\right) \,\frac{dx}{e^x+1}\,  \label{cv2}
\end{equation}
The heat capacity $C_V$ vs $\beta $ is given in Fig.~\ref{h3}. At high
temperature ($\beta \ll 1$) they all tend to (\ref{c0}), while at low
temperature ($\beta \gg 1$) the discrepancy between tachyonic and subluminal
excitations is much more evident.

\section{Discussion}

\begin{figure}[tbp]
\caption{The particle number density of ideal gases of thermal excitations
in the unit of $n_0$ (\ref{n0}) vs temperature variable $\beta = m/T$. Solid
line: tachyonic fermions, dotted line: massive subluminal fermions.}
\label{h0}{\includegraphics[scale=0.6]{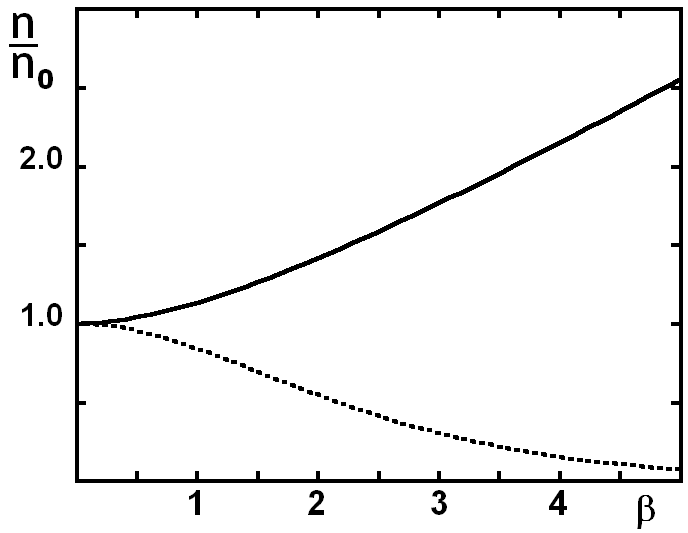}}
\end{figure}

An ideal gas of tachyonic thermal excitations reveals peculiar properties.
Its particle number density (\ref{s2}) is always higher than the particle
number density massless excitations 
\begin{equation}
n_0=\frac{\gamma \zeta \left( 3\right) T^3}{\pi ^2}  \label{n0}
\end{equation}
or the particle number density of massive subluminal excitations at the same
temperature (\ref{s2}), see Fig.~\ref{h0}).

\begin{figure}[tbp]
\caption{The ratio of pressure to the energy density $P/E$ vs $\beta = m/T$.
Solid line: tachyonic fermions, dotted line: massive subluminal fermions.}
\label{h1}{\includegraphics[scale=0.6]{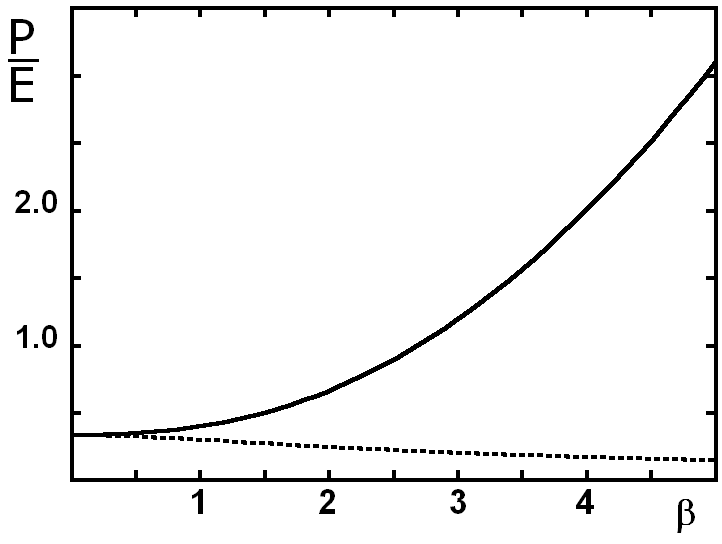}}
\end{figure}

The sound speed (\ref{s2}) in an ideal gas of tachyonic excitations (Fig.~%
\ref{h1}) is superluminal when 
\begin{equation}
\beta >\beta _c\cong 4.36  \label{cr1}
\end{equation}
i.e. when 
\begin{equation}
T<T_c\cong 0.23m  \label{cr2}
\end{equation}
Thus, an ideal Fermi gas of tachyonic thermal excitations satisfies the
causality (\ref{c}) and can form continuous medium when its temperature is
higher the critical value $T_c$ (\ref{cr2}).

While the equation of state of massive subluminal excitations is ''softer''
than the EOS\ of photon gas 
\begin{equation}
P<\frac E3  \label{pho}
\end{equation}
the EOS of tachyonic excitations is always relatively ''stiff'' 
\begin{equation}
P>\frac E3  \label{pho2}
\end{equation}
and when 
\begin{equation}
\beta >\beta _1\cong 2.69m  \label{cr3}
\end{equation}
it becomes ''hyperstiff'' 
\begin{equation}
P>E  \label{hyp}
\end{equation}
see Fig.~\ref{h1}. At the critical temperature $T_c$ (\ref{cr2}) the
tachyonic gas attains minimum possible value of the energy density and
pressure 
\begin{equation}
E_c=1.40\times 10^{-3}\gamma m^4  \label{e44}
\end{equation}

\begin{equation}
P_c=3.32\times 10^{-3}\gamma m^4  \label{p44}
\end{equation}
while 
\begin{equation}
P_c\cong 2.36E_c  \label{hyp2}
\end{equation}
For example, for the nucleon mass $m=m_p=939\,\mathrm{MeV}$ formulas (\ref
{e44})-(\ref{p44}) yield the following estimation 
\begin{equation}
E_c=282\,\mathrm{MeV\cdot fm}^{-3}  \label{e55}
\end{equation}

\begin{equation}
P_c=670\,\mathrm{MeV\cdot fm}^{-3}  \label{p55}
\end{equation}
that is of the same order as the nuclear energy $m_pn_{nm}=158\,\mathrm{%
MeV\cdot fm}^{-3}$ at the saturation density $n_{nm}=0.17\,\mathrm{fm}^{-3}$.

\begin{figure}[tbp]
\caption{Heat capacity $C_V$ in the unit of $C_0$ (\ref{c0}) vs $\beta = m/T$%
. Solid line: tachyonic fermions, dotted line: massive subluminal fermions}
\label{h3}{\includegraphics[scale=0.6]{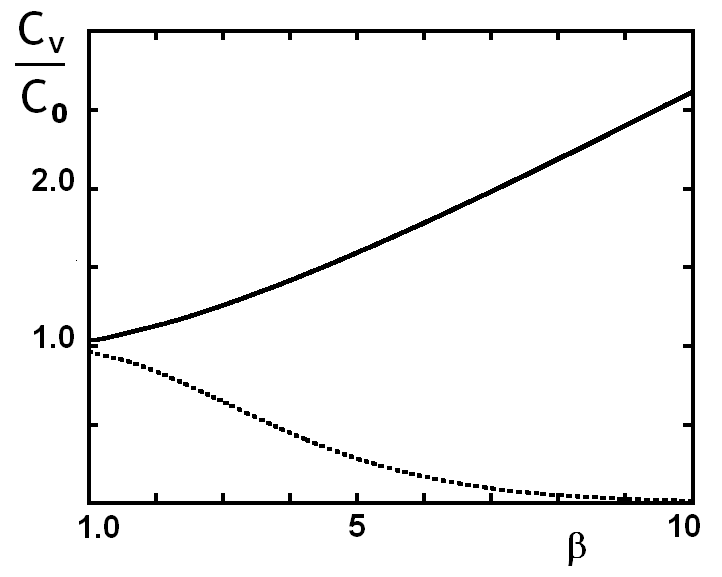}}
\end{figure}

The heat capacity of tachyons is higher than the heat capacity of massless
and massive subluminal fermions (see Fig.~\ref{h3}) attaining its maximum
value 
\begin{equation}
C_V\cong 1.70\gamma T_c^3=1.48C_0  \label{cmx}
\end{equation}
at $T\rightarrow T_c$ and exceeding almost 4 times the heat capacity of
massive subluminal fermions at the same temperature.

The main conclusion of the present study: tachyonic thermal excitations
result in the breaking of the causality (\ref{c}) when the temperature is
below the critical value $T<T_c$ (\ref{cr2}), and their EOS near this point
is ''hyperstiff'' (\ref{hyp}). This may give new ideas for further research
of material medium with tachyonic excitations, where the latter may cause
instability of the whole composite system.

The authors are grateful to Erwin Schmidt for helpful conversations.

\newpage

\end{document}